\documentstyle[amssymb,12pt]{article}

\input{tcilatex}
\begin{document}

\begin{flushright}
{\it PP-IOP-99/24 \\}
\end{flushright}
\vskip 2in

\begin{center}
{\Large {\bf From Prototype SU(5) to Realistic SU(7) SUSY GUT }}

\vspace{30pt}

{\bf J.L. Chkareuli }$^{a}${\bf , C.D. Froggatt }$^{b}${\bf , I.G. Gogoladze 
}$^{a,c}${\bf \ and A.B. Kobakhidze }$^{a,d}${\bf \ } \vspace{6pt}

$^{a}${\em \ }{\small {\it Institute of Physics, Georgian Academy of
Sciences, 380077 Tbilisi, Georgia}}

$^{b}${\small {\it Department of Physics and Astronomy, Glasgow University,
Glasgow G12 8QQ, Scotland}}

$^{c}$ {\small {\it International Centre for Theoretical Physics, 34100
Trieste, Italy }}

$^{d}$ {\em Department of Physics, University of Helsinki, 00014 Helsinki,
Finland\\[0pt]
}

\vskip .5in {\bf Abstract}
\end{center}

We construct a realistic $SU(7)$ model which provides answers to many
questions presently facing the prototype $SU(5)$ SUSY GUT. Among them are a
solution to the doublet-triplet splitting problem, string scale unification,
proton decay, the hierarchy of baryon vs lepton number violation and
neutrino masses.

\thispagestyle{empty} \newpage

\section{Introduction}

Presently, leaving aside the non-supersymmetric Grand Unified Theories which
certainly contradict experiment unless some special extension of the
particle spectrum at intermediate scale(s) is made \cite{1}, even the
commonly accepted $SU(5)$, $SO(10)$ and $E(6)$ SUSY GUTs are far from
perfect. The problems, as they appear for the prototype supersymmetric $%
SU(5) $ model (with a minimal matter, Higgs and gauge boson content) \cite{2}%
, can conventionally be classified as phenomenological and conceptual.

The phenomenological problems include:\newline
{\bf (1)} The large value of the strong coupling $\alpha _{s}(M_{Z})$
predicted, $\alpha _{s}(M_{Z})>0.126$ for the effective SUSY scale $%
M_{SUSY}<1$ TeV \cite{3}, in contrast to the world average value \cite{4} $%
\alpha _{s}(M_{Z})=0.119\pm0.002$;\newline
{\bf (2)} The predicted proton decay rate, due to colour-triplet $H_{c}(%
\stackrel{\_}{H}_{c})$ exchange \cite{2}, is largely excluded by a
combination of the detailed renormalization group (RG) analysis for the
gauge couplings \cite{5} and the improved lower limits on the proton decay
mode $p\to \bar{\nu}K^{+}$ from Super-Kamiokande and on the superparticle
masses from LEP2{\ \cite{6}; } \newline
({\bf 3}) The absence of any sizeable neutrino masses $m_{\nu }\geq 10^{-2}$
eV in the model is in conflict with the atmospheric neutrino deficit data
reported in{\ \cite{SK}}.

Furthermore, the present status of the minimal $SU(5)$ SUSY GUT appears to
be inadequate for the following conceptual reasons:\newline
{\bf (4)} The first one is of course the doublet-triplet splitting problem
for the Weinberg-Salam Higgs field, taken in the fundamental representation
of $SU(5)$, which underlies the gauge hierarchy phenomenon in SUSY GUT \cite
{2}; \newline
{\bf (5)} Then a low unification scale $M_{U}$ is obtained for the MSSM
gauge coupling constants, whose value lies one order of magnitude below that
of the typical string unification scale $M_{str}\simeq 5.3\cdot 10^{17}$ GeV 
\cite{Sstrings};\newline
{\bf (6)} And lastly, the gravitational smearing of its principal
predictions (particularly for $\alpha _{s}(M_{Z})$), due to the
uncontrollable high-dimension operators induced by gravity in the kinetic
terms of the basic SM gauge bosons \cite{hall}, makes the ordinary $SU(5)$
model largely untestable.

The abovementioned three plus three problems seem to be generic for all the
presently popular GUTs. One could imagine that they are all related, in one
way or another, with the still inexplicable gauge hierarchy phenomenon
underlying the Grand Unification of quarks and leptons. It even seems
possible that the true solution to the doublet-triplet problem would itself
specify the complete framework for Grand Unification, including the
fundamental starting gauge symmetry of the GUT involved.

In this connection we would like to call the reader$^{\prime }$s attention
to a novel possibility which opens up \cite{jgk} in some $SU(N)$ GUTs beyond
the prototype $SU(5)$ model: namely, the existence of missing VEV vacuum
configurations for an adjoint scalar, which give a natural solution to the
doublet-triplet problem. According to this missing VEV mechanism, formulated
some time ago \cite{dim}, the basic symmetry breaking adjoint scalar $\Sigma
_{j}^{i}$ ($i,j=1,...,N$) does not develop a VEV in some of the directions
in the $SU(N)$ space and, thereby, splits the masses of a pair of Higgs
fields $H$ and $\overline{H}$ in a hierarchical way through its coupling
with them. The hierarchy is supposed to be such as to give light weak
doublets which break the electroweak gauge symmetry and give masses to up
and down type quarks, on the one hand, and superheavy colour triplets
mediating proton decay, on the other. Interestingly enough, while this
possibility is unrealizable in the standard one-adjoint case \cite{dim}, the
situation radically changes when two adjoint scalar fields are used or,
equally, when the non-renormalisable higher-order $\Sigma $ terms are
included \cite{jgk}. As a result, all the above questions appear to be, at
least partially, answered. We show in Section 2 that the missing VEV
configurations, which ensure the survival of the MSSM at low energies, only
emerge in extended $SU(N)$ GUTs with $N\ge 7$. We consider the minimal $%
SU(7) $ model in detail in the main Section 3 of this paper.

Inasmuch as the missing VEV vacuum must be strictly protected from any large
influence coming from the extra symmetry breaking ($SU(7)\rightarrow SU(5)$%
), we introduce an anomalous $U(1)_{A}$ symmetry, supposedly inherited from
superstrings \cite{gr-sch}, as a custodial symmetry in the model. The
separation of the adjoint scalar and extra symmetry breaking scalar sectors,
provided by the $U(1)_{A}$ symmetry, in fact leads to an increase in the
global symmetry of the model. This global symmetry is partially broken, when
the $SU(7)$ gauge symmetry is spontaneously broken, thus producing a set of
pseudo-Goldstone states of the type

\begin{equation}
5~+~\bar{5}~+~SU(5)\mbox{-}{\rm singlets}  \label{1}
\end{equation}
which gain a mass at the TeV scale due to SUSY breaking. With this
exception, which should be considered as the most generic prediction of the
model, the spectrum at low energies looks just as if one had the prototype $%
SU(5)$ as a starting symmetry. All other $SU(7)$ inherited states, with the
chosen assignment of matter and Higgs superfields, aquire GUT scale masses
during the symmetry breaking, thus completely decoupling from low-energy
physics.

Further, we construct a general $R$-parity violating superpotential where
the effective lepton number violating couplings immediately evolve at the
GUT scale, while the baryon number non-conserving ones are safely projected
out by the missing VEV vacuum configuration involved. Remarkably, the
anomalous $U(1)_{A}$ introduced purely as a missing VEV protecting symmetry
is found to naturally act also as a family symmetry, which can lead to the
observed pattern of quark and lepton masses and mixings.

Another distinctive feature of the $SU(7)$ model considered concerns the
relatively low mass scale $M_{P}^{2/3}M_{SUSY}^{1/3}$ \cite{yan} of the
adjoint moduli fields surviving after $SU(7)$ breaks and, more importantly,
their mass-splitting which inevitably appears in the missing VEV generating
superpotential. This leads to a very different unification picture in $SU(7)$%
. String scale gauge coupling unification in the $SU(7)$ model is explicitly
demonstrated in Section 4, for both small and large $\tan \beta $ values.

Finally, our conclusions are summarized in Section 5.

\section{Motivations for $SU(7)$ GUT}

It is well known that a missing VEV solution for an adjoint scalar $\Sigma $
is absent in the prototype $SU(5)$. Furthermore, even in the general $SU(N)$
GUT, it can not appear in the standard one-adjoint case. The main obstacle
to this is the presence of a cubic term $\Sigma ^{3}$ in the Higgs
superpotential $W$. This cubic term leads to the impracticable trace
condition $Tr\Sigma ^{2}=0$ for the missing VEV vacuum configuration, unless
there is a special fine-tuned cancellation between $Tr\Sigma ^{2}$ and
driving terms stemming from other parts of the superpotential $W$ \cite{dim}.

On its own the elimination of the $\Sigma ^{3}$ term leads to the trivial
unbroken symmetry case. However the inclusion of higher even-order $\Sigma $
terms in the effective superpotential, or the introduction of another
adjoint scalar $\Omega $ with only renormalizable couplings appearing in $W$%
, leads to an all-order missing VEV solution, as was shown in recent papers 
\cite{jgk}.

Let us consider briefly the high-order term case first. The $SU(N)$
invariant superpotential for an adjoint scalar field conditioned also by the
gauge $Z_{2}$ reflection symmetry ($\Sigma \rightarrow -\Sigma $)

\begin{equation}
W_{A}=\frac{1}{2}m\Sigma ^{2}+\frac{\lambda _{1}}{4M_{P}}\Sigma ^{4}+\frac{%
\lambda _{2}}{4M_{P}}\Sigma ^{2}\Sigma ^{2}+...  \label{1b}
\end{equation}
contains, in general, all possible even-order $\Sigma $ terms scaled by
inverse powers of the (conventionally reduced) Planck mass $M_{P}=(8\pi
G_{N})^{-1/2}\simeq 2.4\cdot 10^{18}$ GeV. As one can readily confirm, the
necessary condition for any missing VEV solution to appear in the $%
SU(N)\otimes $ $Z_{2}$ invariant superpotential $W_{A}$ is the tracelessness
of all the odd-order $\Sigma $ terms

\begin{equation}
Tr\Sigma ^{2s+1}=0\mbox{ , } \quad s=0,1,2,...  \label{1c}
\end{equation}
This condition uniquely leads to a missing VEV pattern of the type

\begin{eqnarray}
&&~~~~~~~~\hspace{0.5cm}~~~~\quad N-k\mbox{ \quad \quad }k/2%
\mbox{ \qquad
\quad }k/2  \nonumber \\
&<&\Sigma >=\sigma Diag(\overbrace{~0~...~0~}~,\overbrace{~1~...~1~}~,%
\overbrace{~-1~...-1~})\mbox{,}  \label{1d}
\end{eqnarray}
where the VEV value $\sigma $ is calculated using the $\Sigma $ polynomial
taken in $W_{A}$ (\ref{1b}). The vacuum configuration (\ref{1d}) gives rise
to a particular breaking channel of the starting $SU(N)$ symmetry

\begin{equation}
SU(N)\rightarrow SU(N-k)\otimes SU(k/2)\otimes SU(k/2)\otimes
U(I)_{1}\otimes U(I)_{2}\mbox{ ,}  \label{1e}
\end{equation}
which we will discuss in some detail a little later. For the moment one may
conclude from Eqs. (\ref{1d}, \ref{1e}) that a missing VEV solution, which
retains the ordinary MSSM gauge symmetry $SU(3)_{C}\otimes SU(2)_{W} \otimes
U(1)_{Y}$ at low energies, could actually exist if the $SU(5)$ GUT were
properly extended.

The superpotential (\ref{1b}) could be viewed as an effective one, following
from an ordinary renormalizable two-adjoint superpotential with the second
heavy adjoint scalar integrated out. Hereafter, although both approaches are
closely related, we deal for simplicity with the two-adjoint case. Towards
this end, let us consider in more detail a general $SU(N)$ invariant
renormalizable superpotential for two adjoint scalars $\Sigma $ and $\Omega$%
, satisfying also the gauge-type $Z_{2}$ reflection symmetry ($\Sigma \to
-\Sigma $, $\Omega \to \Omega $) inherited from superstrings 
\begin{equation}
W_{A}=\frac{1}{2}m\Sigma ^{2}+\frac{1}{2}M_{P}\Omega ^{2}+\frac{1}{2}h\Sigma
^{2}\Omega +\frac{1}{3}\lambda \Omega ^{3}.  \label{2}
\end{equation}
Here the second adjoint $\Omega $ can be considered as a state originating
from a massive string mode with the Planck mass $M_{P}$. The basic adjoint $%
\Sigma $ may be taken at another well motivated scale $m\sim
M_{P}^{2/3}M_{SUSY}^{1/3}\sim O(10^{13})$ GeV \cite{yan} where, according to
many string models, the adjoint moduli states $(1_{c},1_{w})$, $%
(1_{c},3_{w}) $ and $(8_{c},1_{w})$ (in a self-evident $SU(3)_{C}\otimes
SU(2)_{W}$ notation) appear. In the present context these states can be
identified as just the non-Goldstone remnants $\Sigma _{0,}$ $\Sigma _{3}$
and $\Sigma _{8} $ of the relatively light adjoint $\Sigma $ which breaks $%
SU(N)$ in some way. However, all our conclusions remain valid for any
reasonable value of $m $, which is the only mass parameter (apart from $%
M_{P} $) in the model considered. In fact we vary $m$ between $10^{12}$ and $%
10^{16}$ GeV.

As a general analysis of the superpotential $W_{A}$ (\ref{2}) shows \cite
{jgk}, there are just four possible VEV patterns for the adjoint scalars $%
\Sigma $ and $\Omega $: (i) the trivial unbroken symmetry case, $\Sigma
=\Omega =0$; (ii) the single-adjoint condensation, $\Sigma =0$, $\Omega \neq
0$; (iii) the $^{\prime \prime }$parallel$^{\prime \prime }$ vacuum
configurations, $\Sigma \propto \Omega $ and (iv) the $^{\prime \prime }$%
orthogonal$^{\prime \prime }$ vacuum configurations, $Tr(\Sigma \Omega )=0$.
The Planck-mass mode $\Omega $, having a cubic term in $W_{A}$, in all
non-trivial cases develops a standard $^{\prime \prime }$single-breaking$%
^{\prime \prime }$ VEV pattern

\begin{eqnarray}
&&~~~~~~~~~~\hspace{0.5cm}~\qquad N-k~~~~~\hspace{0.5cm}~\qquad ~k  \nonumber
\\
&<&\Omega >\mbox{ }=\omega Diag(\overbrace{~1~...~1~}~,~\overbrace{- \frac{%
N-k}{k}...-\frac{N-k}{k}})\mbox{,}  \label{3a}
\end{eqnarray}
which breaks the starting $SU(N)$ symmetry to

\begin{equation}
SU(N)\rightarrow SU(N-k)\otimes SU(k)\otimes U(I)\mbox{ .}  \label{3b}
\end{equation}
However, in case (iv), the basic adjoint $\Sigma $ develops the radically
new missing VEV vacuum configuration (\ref{1d}), thus giving a $^{\prime
\prime }$double breaking$^{\prime \prime }$ of $SU(N)$ to (\ref{1e}). Using
the approximation $\frac{h}{\lambda }>>\frac{m}{M_{P}}$, which is satisfied
for any reasonable values of the couplings $h$ and $\lambda $ in the generic
superpotential $W_{A}$ (\ref{2}), the VEV values are given by

\begin{equation}
\omega =\frac{k}{N-k}\frac{m}{h}\quad \mbox{, }\quad \sigma =\left( \frac{2N%
}{N-k}\right) ^{1/2}\sqrt{mM_{P}}/h  \label{3c}
\end{equation}
respectively. Remarkably, just the light adjoint $\Sigma $ develops the
largest VEV in the model which, for a properly chosen adjoint mass $m$ and
coupling constant $h$, can easily come up to the string scale $M_{str}$ (see
Section 4).

Furthermore, as has already been intimated above, in order to have the
standard gauge symmetry $SU(3)_{C}\otimes SU(2)_{W}\otimes U(1)_{Y}$
remaining after the breaking (\ref{1e}), one must go beyond the prototype $%
SU(5)$. As is easily seen from Eqs. (\ref{1d}, \ref{1e}), there are two
principal possibilities: the weak-component and colour-component missing VEV
solutions respectively. If it is granted that the $^{\prime \prime }$missing
VEV subgroup$^{\prime \prime }$ $SU(N-k)$ in (\ref{1e}) is just the weak
symmetry group $SU(2)_{W}$, as is traditionally argued \cite{dim}, one is
led to $SU(8)$ as the GUT symmetry group ($N-k=2,k/2=3$) \cite{jgk}.
Another, and in fact minimal, possibility is to identify $SU(N-k)$ with the
colour symmetry group $SU(3)_{C}$ in the framework of an $SU(7)$ GUT
symmetry ($N-k=3,$ $k/2=2$); this $SU(7)$ model is considered further here.
The higher $SU(N)$ GUT solutions, if considered, are based on the same two
principal possibilities: the weak-component or colour-component missing VEV
vacuum configurations, respectively.

Let us see now how this missing VEV mechanism works to solve the
doublet-triplet splitting problem in both $SU(8)$ and $SU(7)$ GUTs. It is
supposed that there is a reflection-invariant coupling of the ordinary MSSM
Higgs-boson containing supermultiplets $H$ and $\overline{H}$ with the basic
adjoint $\Sigma $, but not with $\Omega $, in the superpotential $W_{H}$: 
\begin{equation}
W_{H}=f_{0}\overline{H}\Sigma H~~\quad (\Sigma \to -\Sigma ,\mbox{ }%
\overline{H}H\to -\overline{H}H).  \label{4}
\end{equation}
The superfields $H$ and $\overline{H}$ do not develop VEVs during the first
stage of the symmetry breaking. Thus the $W_{H}$ turns into a mass term for $%
H$ and $\overline{H}$ determined by the missing VEV pattern (\ref{1d}). This
vacuum, while giving generally heavy masses (of the order of the GUT scale $%
M_{U}$) to $H$ and $\overline{H}$, leaves their weak components strictly
massless. To be certain of this, we must specify the multiplet structure of $%
H$ and $\overline{H}$ for both the weak-component and the colour-component
missing VEV vacuum configurations, that is in $SU(8)$ and $SU(7)$ GUTs
respectively. In the $SU(8)$ case, $H$ and $\overline{H}$ are fundamental
octets whose weak components (ordinary Higgs doublets) do not get masses
from the basic coupling (\ref{4}). In the $SU(7)$ case, $H$ and $\overline{H}
$ are 2-index antisymmetric $21$-plets which (after projecting out the extra
heavy states, see Section 3.3) contain just a pair of massless Higgs
doublets. Thus, there certainly is a natural doublet-triplet splitting in
both cases and we also have a vanishing $\mu $ term at this stage. However,
one can readily show that the right order $\mu $ term always appears as a
result of radiative corrections at the next stage when SUSY breaks \cite{jgk}%
.

We consider below the minimal $SU(7)$ GUT in some detail.

\section{Basics of the $SU(7)$ model}

\subsection{Matter and Higgs superfields}

By analogy with the standard $SU(5)$ model, we consider the simplest
anomaly--free combination of the fundamental and antisymmetric 2-index
representations of the $SU(7)$ gauge group

\begin{equation}
\left[ \overline{\xi }^{A}+\overline{\zeta }^{A}+\overline{\Psi }^{A}+\Psi
_{[AB]}\right] _{i}  \label{7}
\end{equation}
(where $A,B=1,...,7$ are $SU(7)$ indices) for each of the three
quark--lepton families or generations ($i=1,2,3$). The quark-lepton states
reside in the multiplets $\overline{\Psi }^{A}(\overline{7})+\Psi
_{[AB]}(21) $, while the extra fundamental multiplets $\overline{\xi }^{A}(%
\overline{7})$ and $\overline{\zeta }^{A}(\overline{7})$ are specially
introduced in (\ref{7}) for anomaly cancellation. There is also a set of
Higgs superfields, among which are the two already mentioned adjoint Higgs
multiplets $\Sigma _{B}^{A}(48)$ and $\Omega _{B}^{A}(48)$. They are
responsible here for the GUT scale breaking of $SU(7)$

\begin{equation}
SU(7)\rightarrow SU(3)_{C}\otimes SU(2)_{W}\otimes SU(2)_{E}\otimes
U(I)_{1}\otimes U(I)_{2}  \label{3d}
\end{equation}
according to the VEVs (\ref{1d}, \ref{3a}, \ref{3c}), which now become

\begin{eqnarray}
&<&\Sigma >\mbox{ }=\sigma Diag[0,0,0,1,1,-1,-1]  \label{10} \\
&<&\Omega >\mbox{ }=\omega Diag[1,1,1,-\frac{3}{4},-\frac{3}{4},-\frac{3}{4}%
,-\frac{3}{4}]  \label{10x}
\end{eqnarray}
where $\sigma =\sqrt{14mM_{P}/3}/h,$ $\omega =4m/3h$ and the colour ($C$),
weak ($W$) and extra ($E$) components are indicated. In addition there are a
pair of Higgs multiplets $H_{[AB]}(21)$ and $\overline{H}^{[AB]}(\overline{21%
})$ in conjugate representations, which contain the ordinary electroweak
doublets. One can easily check that, due to their basic coupling (\ref{4})
with the adjoint $\Sigma $, which develops the missing VEV configuration (%
\ref{10}), all the states in the multiplets $H_{[AB]}$ and $\overline{H}%
^{[AB]}$ become superheavy with mass of order $M_{U}$, except for one pair
of colour triplets and two pairs of weak doublets. Thus, in order to have
just the standard pair of MSSM electroweak doublets, the other pair of weak
doublets should be projected out from the massless state spectrum together
with the colour triplet states. This is accomplished by the mixing of $%
H_{[AB]}$ and $\overline{H}^{[AB]}$ with specially introduced heavy scalar
supermultiplets, the 3-index antisymmetric $35$-plets of $SU(7)$ $\Phi
_{[ABC]}$ and $\overline{\Phi }^{[ABC]}$, which contain just the required
states (see Section 3.3). And, finally, there are two fundamental scalar
superfields $\varphi _{A}(7)$ and $\eta _{A}(7)$ and their $^{\prime \prime
} $conjugates$^{\prime \prime }$ $\overline{\varphi }^{A}(\overline{7})$ and 
$\overline{\eta }^{A}(\overline{7})$ which break the extra symmetry at the
GUT scale. We consider this key question first.

\subsection{Extra symmetry breaking}

Inasmuch as the extra symmetry should also be broken 
\begin{equation}
SU(7)\rightarrow SU(5)  \label{5}
\end{equation}
at the GUT scale, in order not to spoil gauge coupling unification, a
question arises: how can the adjoint $\Sigma $ missing VEV configuration (%
\ref{10}) survive so as to be subjected to at most a shift of order the
electroweak scale? This requires, in general, that the superpotential $W_{A}$
(\ref{2}) be strictly protected from any large influence of the scalars $%
\varphi $ and $\eta $, which provide the extra symmetry breaking (\ref{5}).
Technically, such a custodial symmetry may be a superstring-inherited
anomalous $U(1)_{A}$ \cite{gr-sch}, which induces a high-scale extra
symmetry breaking (\ref{5}) through the Fayet-Iliopoulos (FI) $D$-term \cite
{witten}:

\begin{equation}
D_{A}=\xi +\sum Q_{A}^{n}\mid <{\cal F}^{n}>\mid ^{2}\mbox{, \qquad }\xi =%
\frac{TrQ_{A}}{192\pi ^{2}}\mbox{ }g_{str}^{2}M_{P}^{2}\mbox{. \qquad }
\label{6'}
\end{equation}
Here the sum runs over all $^{\prime \prime }$charged$^{\prime \prime }$
scalar fields in the theory, including those which do not develop VEVs and
only contribute to $TrQ_{A}$. For realistic or semi-realistic models, $%
TrQ_{A}$ has turned out to be quite large, $TrQ_{A}=O(100)$ (see the recent
discussion in \cite{zz}). So, the spontaneous breaking scale of the $%
U(1)_{A} $ symmetry and the related extra symmetry (\ref{5}) is naturally
located at the string scale. The protecting anomalous $U(1)_{A}$ symmetry is
supposed to keep the scalars $\varphi $ and $\eta $ essentially decoupled
from the basic adjoint superpotential $W_{A}$ (\ref{2}), so as not to
strongly influence its missing VEV vacuum configuration (\ref{10}) through
the appearance of potentially dangerous couplings of the type $\overline{%
\varphi }\Sigma \varphi $ and $\overline{\eta }\Sigma \eta $.

Anyway, once a separation of the adjoint scalar and extra symmetry breaking
scalar sectors takes place in the supersymmetric $SU(7)\otimes $ $U(1)_{A}$
theory considered, an accidental global symmetry $SU(7)_{\Sigma -\Omega
}\otimes U(7)_{\varphi -\eta }$ appears. This global symmetry is radiatively
broken and one or two families of pseudo-Goldstone (PG) states of type (\ref
{1}) are produced at a TeV scale, where SUSY softly breaks \cite{jgk}. The
two-family case corresponds to the most degenerate Higgs potential, where
the scalars $\varphi $ and $\eta $ are only allowed to appear through the
basic $SU(7)$ and $U(1)_{A}$ $D$-terms and, thereby, increase their global
symmetry to $U(7)_{\varphi }\otimes U(7)_{\eta }$. This two-family case
occurs when the $U(1)_{A}$ charges of the bilinears $\overline{\varphi }%
\varphi ,$ $\overline{\eta }\eta ,$ $\overline{\varphi }\eta $ and $%
\overline{\eta }\varphi $ are all positive (or all negative), so that they
can not appear in the $SU(7)\otimes U(1)_{A}$ invariant superpotential in
any order.

However, remarkably, it is possible for the adjoint and fundamental scalar
sectors in the superpotential to overlap without disturbing the adjoint
missing VEV configuration. This naturally occurs when the scalars $\varphi $
and $\eta $ are conditioned by the $U(1)_{A}$ symmetry to develop orthogonal
VEVs along the $^{\prime \prime }$extra$^{\prime \prime }$ directions 
\begin{equation}
\varphi _{A}=\delta _{A6}V_{1,}\mbox{ }\qquad \eta _{A}=\delta _{A7}V_{2}
\label{6''}
\end{equation}
The simplest choice of such safe mixing terms is given by dimension-5
operators, invariant under the reflection symmetry $\Sigma \rightarrow
-\Sigma $, $\overline{\varphi }\rightarrow -\overline{\varphi }$, $\overline{%
\eta }\rightarrow -\overline{\eta }$, of the type\footnote{%
Enlarging the scalar sector properly one can write a renormalizable
superpotential as well.}

\begin{equation}
W_{H1}=\frac{1}{M_{P}}\overline{\eta }[a\cdot S\Sigma +b\cdot \overline{%
\varphi }\varphi ]\varphi \mbox{ }  \label{6'''}
\end{equation}
The dimensionless coupling constants $a$ and $b$ are both of order $O(1)$
and $S$ is some new singlet superfield which gets its VEV through the FI $D$%
-term (\ref{6'}), just as the scalars $\varphi $ and $\eta $ do. One can
consider the field $S$ as a basic carrier of unit $U(1)_{A}$ charge in the
model. In terms of its charge, the charges of the bilinears in $W_{H1}$ are
determined to be $+1$ for $\overline{\varphi }\varphi $ and $-1$ for $%
\overline{\eta }\varphi $, while the charges of the bilinears $\overline{%
\eta }\eta $ and $\overline{\varphi }\eta $ are not yet determined from the
couplings (\ref{6'''}). The latter charges can be taken to be positive, as
follows when we consider the other coupling terms in Sections 3.3 - 3.6.
This implies that any terms containing $\varphi $ and $\eta $ scalars in the
superpotential must also include the bilinear $\overline{\eta }\varphi $, so
as to properly compensate the $U(1)_{A}$ charges. However, for a vacuum
configuration where the orthogonality condition $\overline{\eta }\varphi =0$
naturally arises, this gives an all-order solution excluding the dangerous $%
\overline{\varphi }\Sigma \varphi $ and $\overline{\eta }\Sigma \eta $
terms. In fact this orthogonality condition is precisely one of the
conditions satisfied at the SUSY invariant global minimum of the Higgs
potential, as follows from the vanishing $F$-terms of the superfields
involved in (\ref{6'''}):

\begin{equation}
\overline{\eta }\varphi =0,\mbox{ \quad }\overline{\varphi }\varphi =\frac{a%
}{b}\cdot S\sigma \mbox{ }  \label{7a}
\end{equation}
Here the VEV value, appearing on the extra symmetry components of the
adjoint $\Sigma $ (\ref{10}), has been used. One can now readily see that a
non-diagonal mass-term appears for the PG states related with the multiplets 
$\varphi $ and $\eta $ and their $^{\prime \prime }$conjugates$^{\prime
\prime }$

\begin{equation}
M_{\overline{\eta }\varphi }\equiv [W_{H1}^{^{\prime \prime }}]_{\overline{%
\eta }\varphi }=\frac{a}{M_{P}}\cdot S(\Sigma +\sigma I)  \label{7b}
\end{equation}
where $I$ is the $7\times 7$ unit matrix. This explicitly shows that one
superposition of the two PG $5~+~\bar{5}$ families (\ref{1}) becomes heavy,
while the other is always left massless. In fact this result is a general
consequence of the symmetry breaking pattern involved. The point is that
neither of the other mass-terms $M_{\overline{\varphi }\eta }$, $M_{%
\overline{\varphi }\varphi }$ and $M_{\overline{\eta }\eta }$ can be allowed
by the $U(1)_{A}$ symmetry for any generalization of the superpotential $%
W_{H1}$ (\ref{6'''}); otherwise the dangerous $\overline{\varphi }\Sigma
\varphi $ and $\overline{\eta }\Sigma \eta $ couplings inevitably appear as
well. Similarly, in the general $SU(N)$ case, it can be shown \cite{CF} that
at least one family of PG states of type (\ref{1}) always exists. Together
with the ordinary quarks and leptons and their superpartners these PG
states, both bosons and fermions, determine the particle spectrum at low
energies. In most of what follows the existence of just one family of PG
states at the sub-TeV scale will be assumed.

\subsection{Heavy states}

We demonstrate below that, when the extra symmetry is broken (\ref{5}), all
the states in the $SU(7)$ model, beyond the ordinary MSSM particle spectrum
plus one family of pseudogoldstones (\ref{1}), acquire masses of order the
GUT scale. An exception can be made for the sterile states (the states in
the matter multiplets (\ref{7}) having the extra symmetry charges only)
whose fermionic components might be referred to as sterile neutrinos (see
Section 3.4).

First of all let us consider the Higgs sector and show that all the states
in the basic Higgs multiplets $H_{[AB]}$ and $\overline{H}^{[AB]}$ become
superheavy, except for one pair of weak doublets. One can readily check
that, when the colour-component missing VEV solution (\ref{10}) is
substituted into the superpotential $W_H$ (\ref{4}), superheavy masses are
generated for most of the components of the $H$ and $\overline{H}$
multiplets. However, the following states (weak, colour and extra symmetry
components are explicitly indicated)

\begin{equation}
H_{w6\mbox{ }}\mbox{, \quad }\overline{H}_{\mbox{ }}^{w6}\mbox{, \quad }H_{w7%
\mbox{ }}\mbox{, \quad }\overline{H}^{w7}\mbox{, \quad }H_{[cc^{\prime }]%
\mbox{ }}\mbox{, \quad }\overline{H}^{[cc^{\prime }]}  \label{23a}
\end{equation}
remain massless at this stage of $SU(7)\ $symmetry breaking. Therefore one
of the two pairs of weak doublets in (\ref{23a}), as well as the colour
triplets, must also become heavy in order to obtain the ordinary picture of
MSSM at low energies. This happens as a result of mixing $H$ and $\overline{H%
}$ with the specially introduced (see Section 3.1) superheavy scalar
supermultiplets $\Phi _{[ABC]}$ and $\overline{\Phi }^{[ABC]}$ in the basic
Higgs superpotential

\begin{equation}
W_{H2}=f\cdot H\overline{\Phi }\varphi +\overline{f}\cdot \overline{H}\Phi 
\overline{\varphi }+y\cdot S\overline{\Phi }\Phi \mbox{.}  \label{www}
\end{equation}
Here $f$, $\overline{f}$ and $y$ are dimensionless coupling constants. When
the scalars $\varphi $ and $\eta $ get their VEVs, thus breaking the extra
symmetry, the required mixing terms are generated. It is worth noting that
the presence of the $^{\prime \prime }$conjugated$^{\prime \prime }$ $%
\overline{\Phi }-H\ $ and $\Phi -\overline{H}$ mixings in $W_{H2}$ could
allow the dangerous $\overline{\varphi }\Sigma \varphi $ and $\overline{\eta 
}\Sigma \eta $ terms, destroying the missing VEV solution, unless the
bilinear term$\ \overline{\Phi }\Phi $ has a nonzero $U(1)_{A}$ charge.
Therefore, this term appears in $W_{H2}$ together with the singlet scalar
superfield $S$ -- the basic $U(1)_{A}$ charge carrier introduced earlier in $%
W_{H1}$ (\ref{6'''}).

It is easy to see now that the $W_{H2}$ couplings (\ref{www}) will rearrange
the mass spectrum of the states (\ref{23a}), so as to leave just a standard
pair of MSSM electroweak doublets massless. Considering the mixing of the
colour triplet states first, one can see from the $2\times 2$ mass matrix
for the states $H_{[cc^{\prime }]\mbox{ }}$ and $\overline{H}^{[cc^{\prime
}]}$ and the double-coloured components $\Phi _{[cc^{\prime }6]}$ and $%
\overline{\Phi }^{[cc^{\prime }6]}$ that, when properly diagonalized, the
colour components in (\ref{23a}) obtain a mass $M_{*}$ of order 
\begin{equation}
M_{*}\sim \frac{f\overline{f\mbox{ }}}{y}\frac{<\varphi ><\overline{\varphi }%
>}{S}\sim M_{U}.  \label{23b}
\end{equation}
The combination of primary coupling constants $f,$ $\overline{f\mbox{ }}$
and $y$ in (\ref{23b}), can be taken $O(1)$ in general. For the weak doublet
case, there is a $3\times 3$ mass matrix corresponding to the mixing of the
states $H_{w6\mbox{ }}$, $H_{w7\mbox{ }}$ and $\Phi _{[w67]}$, and their $%
^{\prime \prime }$conjugates$^{\prime \prime }$ $\overline{H}^{w6}$, $%
\overline{H}^{w7}$ and $\overline{\Phi }^{[w67]}$ respectively. After
diagonalization this matrix leaves, as can readily be checked, just one pair
of weak-doublets $H_{w6}$ and $\overline{H}^{w6}$ strictly massless, while
the other pair $H_{w7\mbox{ }}$ and $\overline{H}^{w7}$ acquires a mass of
order $M_{*}$ (\ref{23b}).

It seems reasonable to assume that the components of $H$ and $\overline{H}$,
which get masses from their direct coupling (\ref{4}) with the basic adjoint 
$\Sigma $, should have a mass of order $M_{U}$, while the states in (\ref
{23a}), which develop a mass $M_{*}$ from their mixing with the heavy $\Phi $
and $\overline{\Phi }$ multiplets (\ref{www}), could naturally be relatively
light, $M_{*}=O(10^{-2}\div 1)M_{U}$. In this case, the proton decay
inducing colour-triplet states $H_{c6}$ and $\overline{H}^{c6}$, which are
partners of the light weak-doublets $H_{w6}$ and $\overline{H}^{w6}$, are
located at the GUT scale $M_{U}$, while the double-coloured states $%
H_{[cc^{\prime }]\mbox{ }}$and\quad $\overline{H}^{[cc^{\prime }]}$ (\ref
{23a}), which can not induce proton decay, are relatively light. Thus the
problem of an unacceptably fast proton decay, due to dimension-5 operators,
would be naturally solved in the $SU(7)$ model considered. At the same time
the double-coloured states $H_{[cc^{\prime }]\mbox{ }}$ and $\overline{H}%
^{[cc^{\prime }]}$, if taken relatively light, would properly contribute to
the running of the gauge coupling constants (see Section 4).

Next we consider the additional matter multiplets and, two anti-septets $%
\overline{\xi }^{A}$ and $\overline{\zeta }^{A}$ for each of the three
generations, which were introduced (Section 3.1) for $SU(7)$ anomaly
cancellation. They are assumed to form their masses by combining with the
basic 2-index multiplet $\Psi _{[AB]}$ of their own generation

\begin{equation}
W_{Y1}=g_{_{\xi }}\cdot \overline{\xi }^{A}\Psi _{[A6]}\overline{\varphi }%
^{6}+g_{_{\zeta }}\cdot \overline{\zeta }^{A}\Psi _{[A7]}\overline{\eta }%
^{7}.  \label{23c}
\end{equation}
Here the $SU(7)$ index $A$ and the extra symmetry subgroup indices $6$ and $%
7 $, along which the VEVs (\ref{6''}) develop, are explicitly indicated ($%
g_{_{\xi ,\mbox{ }\zeta }}$ are coupling constants). So, again one can see
that any state in the multiplets $\overline{\xi }$ and $\overline{\zeta }$
(for all three generations) carrying colour and/or electric charge acquires
a mass of order $M_{U}$.

\subsection{Sterile neutrinos}

The states in the matter multiplets (\ref{7}) of each generation, which only
have charges under the extra symmetry

\begin{equation}
\Psi _{[67]},\mbox{ \quad }\overline{\Psi }^{6},\mbox{ \quad }\overline{\Psi 
}^{7},\mbox{ \quad }\overline{\xi }^{6},\mbox{ \quad }\overline{\xi }^{7},%
\mbox{ \quad }\overline{\zeta }^{6},\mbox{ \quad }\overline{\zeta }^{7}
\label{23d}
\end{equation}
are of particular interest as possible $SU(7)$ candidates for sterile
neutrinos. One can see that two superpositions of $\Psi _{[67]}$, $\overline{%
\xi }^{7}$ and $\overline{\zeta }^{6}$ acquire masses of order $<\overline{%
\varphi }>$ $\sim $ $<\overline{\eta }>$ $\sim M_{U}$ from the couplings (%
\ref{23c}). As to the other states in (\ref{23d}), their masses depend on
the $U(1)_{A}$ charges assigned to the matter and Higgs superfields
involved. If one takes the simple set of charges presented in Table 1, all
of them in fact can get masses from the allowed high-order terms (inherited
from superstrings or induced by gravitational corrections) of the type

\begin{eqnarray}
M_{P}W_{Y2} &=&[a_{1}\overline{\Psi }\overline{\Psi }+a_{2}\overline{\Psi }%
\overline{\zeta }\frac{S}{M_{P}}+a_{3}\overline{\zeta }\overline{\zeta }(%
\frac{S}{M_{P}})^{2}+a_{4}\overline{\Psi }\overline{\xi }(\frac{S}{M_{P}}%
)^{3}]\cdot \varphi \eta +  \nonumber \\
&&[a_{5}\overline{\zeta }\overline{\xi }+a_{6}\overline{\xi }\overline{\xi }(%
\frac{S}{M_{P}})^{2}]\cdot \eta \eta +...  \label{23e}
\end{eqnarray}
where the higher order terms indicated by dots are ignored (overall $SU(7)$
index contraction is implied and the dimensionless coupling constants $a_{1}$%
,..., $a_{6}$ are assumed to be all of $O(1)$). It follows, from the
couplings in $W_{Y1}$ (\ref{23c}) and $W_{Y2}$ (\ref{23e}), that for each
generation the symmetric $7\times 7$ mass matrix of the sterile states (\ref
{23d}) has all its eigenvalues non-zero in general. The smallest eigenvalue
is just given by the order of the highest term kept in the superpotential $%
W_{Y2}$ and is thus of order $\frac{M_{U}^{5}}{M_{P}^{4}}$.

Some of these sterile states should be considered as candidates for
right-handed neutrinos. The traditionally used Planck or GUT mass scale
right-handed neutrino leads, via the well-known see-saw mechanism \cite
{s-saw}, to ordinary neutrino masses in the range $m_{\nu }=10^{-5}\div
10^{-3}$ eV or lower, which cannot explain the recent SuperKamiokande
atmospheric neutrino data {\cite{SK}}. By contrast, the sterile states
discussed here\footnote{%
Other potentially interesting right-handed neutrino candidates, also well
motivated in the $SU(7)$ model, are the fermionic superpartners of the
adjoint moduli states $\Sigma _{0}$ and $\Sigma _{3}$ at a scale of $%
M_{P}^{2/3}M_{SUSY}^{1/3}$ stemming from superstrings {\cite{yan}}.} could
naturally lead to neutrino mass(es) in just the required region $m_{\nu
}=10^{-2}\div 1$ eV.

Interestingly, with the choice of $U(1)_{A}$ charges taken in Table 1, the
direct Higgs-matter mixing terms such as

\begin{equation}
W_{Y3}=b_{1}\overline{\xi }\eta S+b_{2}\overline{\Psi }\varphi \frac{S^{2}}{%
M_{P}}+b_{3}\overline{\zeta }\varphi \frac{S^{3}}{M_{P}^{2}}+...
\label{23ee}
\end{equation}
can also evolve, provided that $R$-parity symmetry is not assumed. They will
induce the condensation of some of the heavy sterile sneutrinos, just like
takes place for ordinary sneutrinos in the bilinear Higgs-lepton mixing
model \cite{lnv}.

At the same time, it is well to bear in mind that some other choice of the $%
U(1)_{A}$ charges, different from those in Table 1, could (partially or
completely) forbid the couplings (\ref{23e}) and (\ref{23ee}), thus leading
to certain light and even strictly massless sterile neutrinos. As is well
known, they are not ruled out by experiment {\cite{SK},\cite{lnv}.}

\subsection{Lepton number violation}

One would think that there is no fundamental reason for exact $R$-parity ($%
RP $) symmetry in the framework of supersymmetric GUTs, where not only
fermions but also their scalar superpartners are the natural carriers of
lepton and baryon numbers. Accordingly, we suppose that all the generalized
Yukawa couplings, the $RP$-conserving (ordinary up and down fermion
Yukawas), as well as the $RP$-violating ones allowed by $SU(7)\otimes
U(1)_{A}$ symmetry, are given by a similar set of dimension-5 operators. We
note that the usual dimension-4 trilinear Yukawa couplings are forbidden by
the underlying gauge invariance. The dimension-5 operators take the form ($%
i,j,k=1,2,3$ are the generation indices) 
\begin{equation}
{\cal O}_{ij}^{up}=\frac{G_{ij}^{u}}{M_{P}}\left( \Psi _{i}\Psi _{j}\right)
(H\eta )  \label{14}
\end{equation}
\begin{equation}
{\cal O}_{ij}^{down}=\frac{G_{ij}^{d}}{M_{P}}(\overline{\Psi }_{i}\Psi _{j})(%
\overline{H}\varphi )  \label{15}
\end{equation}
\begin{equation}
{\cal O}_{ijk}^{rpv}=\frac{G_{ijk}}{M_{P}}(\overline{\Psi }_{i}\Psi _{j})(%
\overline{\Psi }_{k}\Sigma )  \label{16}
\end{equation}
and are collected in the Yukawa part of the superpotential $W_{Y}$. As
specified above (Section 3.1), each of the three generations of quarks and
leptons lies in two multiplets $\overline{\Psi }^{A}+\Psi _{[AB]}$ of $SU(7)$%
. Also the ordinary electroweak doublets are contained in the multiplets $%
H_{[AB]}$ and $\overline{H}^{[AB]}$, while the scalars $\varphi $ and $\eta $
and their $^{\prime \prime }$conjugates$^{\prime \prime }$, which break $%
SU(7)$ to $SU(5)$, are fundamental septets and anti-septets, respectively.

Now, on substituting VEVs for the scalars $\Sigma $ (\ref{10}), $\varphi $
and $\eta $ (\ref{6''}) in the basic operators (\ref{14}--\ref{16}), one
obtains at low energies the effective renormalisable Yukawa and lepton
number violating (LNV) interactions with coupling constants

\begin{equation}
Y_{ij}^{u}=G_{ij}^{u}\frac{<\eta >}{M_{P}}\mbox{ , \quad }%
Y_{ij}^{d}=G_{ij}^{d}\frac{<\varphi >}{M_{P}}\mbox{ , \quad }\Lambda
_{ijk}=G_{ijk}\frac{<\Sigma >}{M_{P}}\mbox{ .}  \label{Y}
\end{equation}
At the same time the baryon number violating (BNV) couplings prove to be
completely eliminated. The key element here turns out to be that the adjoint
field $\Sigma $, involved in the effective couplings (\ref{16}), develops a
VEV configuration with strictly zero colour components (\ref{10}) in the
SUSY limit. However, at the next stage when SUSY breaks, the radiative
corrections will shift the missing VEV components of $\Sigma $ to some
nonzero values of order $M_{SUSY}$. In this way the ordinary $\mu $-term of
the MSSM, on the one hand, and baryon number violating couplings with
hierarchically small coupling constants of the order $M_{SUSY}/M_{U}$, on
the other, are induced. So, our missing VEV solution to the gauge hierarchy
problem leads in fact to a similar hierarchy of baryon vs lepton number
violation.

The ${\cal O}$-operators (\ref{14}--\ref{16}) can be viewed as effective
interactions generated through the exchange of some superheavy states, which
we can interpret as massive string modes. When they are generated by the
exchange of the same superheavy multiplet (formed from a pair of fundamental
septets $7+\overline{7}$), the operators (\ref{15}) and (\ref{16}) appear
with the dimensionless effective coupling constants aligned in flavour space 
\cite{JClnv}:

\begin{equation}
\Lambda _{ijk}=Y_{ij}^{d}\cdot \epsilon _{k}^{\ }.  \label{22a}
\end{equation}
Here the parameters $\epsilon _{k}^{\ }$ ($k=1,2,3)$ include some known
combination of the primary dimensionless coupling constants and a ratio of
the VEVs of the scalars $\Sigma $ and $\varphi $. This strict relation,
between general $RP$-violating and down fermion Yukawa coupling constants,
is then split into the ones for charged lepton ($cl$) and down quark ($dq)$
LNV couplings respectively,

\begin{equation}
\lambda _{ijk}=Y_{ij}^{cl}\cdot \epsilon _{k}\mbox{ ,\qquad }\lambda
_{ijk}^{\prime }=Y_{ij}^{dq}\cdot \epsilon _{k},  \label{23}
\end{equation}
when it is evolved from the GUT scale down to low energies.

Therefore, the postulated common origin of all the generalized Yukawa
couplings, both $RP$-conserving and $RP$-violating, at the GUT scale results
in some minimal form of lepton number violation, with the proviso that
appropriate mediating superheavy-states exist. As a result, all significant
physical manifestations of LNV reduce to those of the effective trilinear
couplings\footnote{%
Here $L$ and $Q$ denote lepton and quark $SU(2)$ doublet superfields, while $%
\overline{E}$ and $\overline{D}$ denote lepton and down quark $SU(2)$
singlet superfields.} $LL\overline{E}$ and $LQ\overline{D}$ aligned, both in
magnitude and orientation in flavour space, with the down fermion (charged
lepton and down quark) effective Yukawa couplings. However the effective
bilinear terms $\mu _{i}L_{i}H$ appear to be generically suppressed by the
custodial $U(1)_{A}$ symmetry involved. Detailed phenomenology of this model
related to the flavor-changing processes both in quark and lepton sectors,
radiatively induced neutrino masses and decays of the LSP can be found in a
recent paper \cite{JClnv}.

\subsection{Masses and mixings of quarks and leptons}

So far we have used an anomalous $U(1)_{A}$ symmetry purely as a missing VEV
protecting symmetry in the $SU(7)$ model. In Table \ref{1} we list the $%
U(1)_{A}$ ($e^{iQ_{A}\theta }$) and $Z_{2}$ ($e^{in_{Z}\pi }$) charges,
which allow just the couplings appearing in the total superpotential $W_{T}$:

\begin{equation}
W_{T}=W_{A}+W_{H}+W_{H1}+W_{H2}+W_{Y}+W_{Y1}+W_{Y2}+W_{Y3}  \label{wwww}
\end{equation}
(see Eqs.~(\ref{2}, \ref{4}, \ref{6'''}, \ref{www}, \ref{23c}, \ref{23e}-\ref
{16})). \medskip The superpotential $W_{T}$ in turn completely fixes all the 
$Q_{A}$ charges in terms of the charge of the singlet scalar superfield $S$,
which we take to be unity $Q_{A}^{S}=1$. Remarkably, as one can see from
Table \ref{1}, the $Q_{A}$ charges of the scalar bilinears $\overline{%
\varphi }\varphi $ and $\overline{\eta }\eta $, which break the extra
symmetry ($SU(7)\rightarrow SU(5)$), happen to be positive, while the
adjoint scalar $\Sigma $ has $Q_{A}=0$. Hence the potentially dangerous $%
\overline{\varphi }\Sigma \varphi $ and $\overline{\eta }\Sigma \eta $
couplings are forbidden, thus providing an all-order solution to the
doublet-triplet splitting problem in the $SU(7)$ model.

However, together with its protecting function, it is tempting to treat the
anomalous $U(1)_{A}$ also as a family symmetry \cite{IR}. Then the zero
charges $Q_{A}$ presented in Table \ref{1} for the matter multiplets $\Psi
_{[AB]}$ and $\overline{\Psi }^{A}$ would only be assigned to the third
family of quarks and leptons, while the other two families would be assigned
some nonzero $Q_{A}$ charges. The observed hierarchy of quark-lepton masses
and mixings might then be generated via the Froggatt-Nielsen mechanism \cite
{colin}. In the case at hand this hierarchy is described by the natural
expansion parameter $\epsilon =\frac{<S>}{M_{P}}\sim \frac{M_{str}}{M_{P}}%
(\simeq 0.2)$. One can readily see that a simple choice of $Q_{A}$ charges
for the matter multiplets $\Psi _{[AB]i}$ and $\overline{\Psi }_{i}^{A}$ (as
explicitly indicated in brackets in units of $Q_{A}^{S}$ ) leads to Yukawa
matrices of the type

\begin{equation}
\begin{array}{ccc}
& {\ 
\begin{array}{ccc}
\hspace{-5mm}~\Psi _{1}^{(-3)} & \,\,\Psi _{2}^{(-2)} & \,\,\Psi _{3}^{(0)}
\end{array}
} &  \\ 
\vspace{2mm}{Y}^{u}\propto 
\begin{array}{c}
\Psi _{1}^{(-3)} \\ 
\Psi _{2}^{(-2)} \\ 
\Psi _{3}^{(0)}
\end{array}
\!\!\!\!\! & {\left( 
\begin{array}{ccc}
\,\,\epsilon ^{6}~~ & \,\,\epsilon ^{5}~~ & \,\,\epsilon ^{3} \\ 
\,\,\epsilon ^{5}~~ & \,\,\epsilon ^{4}~~ & \,\,\epsilon ^{2} \\ 
\,\,\epsilon ^{3}~~ & \,\,\epsilon ^{2}~~ & \,\,1
\end{array}
\right) }~ & 
\end{array}
\!\!~~~~~  \label{23f}
\end{equation}
and

\begin{equation}
\begin{array}{ccc}
& {\ 
\begin{array}{ccc}
\hspace{-5mm}~\Psi _{1}^{(-3)} & \,\,\Psi _{2}^{(-2)} & \,\,\Psi _{3}^{(0)}
\end{array}
} &  \\ 
\vspace{2mm}{Y}^{d}\propto 
\begin{array}{c}
\overline{\Psi }_{1}^{(-2)} \\ 
\overline{\Psi }_{2}^{(0)} \\ 
\overline{\Psi }_{3}^{(0)}
\end{array}
\!\!\!\!\! & \mbox{ }{\left( 
\begin{array}{ccc}
\,\,\epsilon ^{5}~~ & \,\,\epsilon ^{4}~~ & \,\,\epsilon ^{2} \\ 
\,\,\epsilon ^{3}~~ & \,\,\epsilon ^{2}~~ & \,\,1 \\ 
\,\,\epsilon ^{3}~~ & \,\,\epsilon ^{2}~~ & \,\,1
\end{array}
\right) }~. & 
\end{array}
\!\!~~~~~  \label{23g}
\end{equation}
These matrices yield a quark and charged lepton mass hierarchy similar to
the observed pattern, as well as the right orders of magnitude for the CKM
matrix elements (see also \cite{shafi}). Phenomenologically, in the case of $%
Y^{u}$ (\ref{23f}) the constant of proportionality is of order unity.
However, in the case of $Y^{d}$ (\ref{23g}), the constant of proportionality
is only of order unity for large values of $\tan \beta $. In the case of
small $\tan \beta $ it is necessary to assume that the underlying
fundamental couplings give a constant of proportionality of order $0.01$%
\footnote{%
It must be admitted that the origin of this small factor is not understood,
just as the origin of a large value of $\tan \beta $ is not understood.
Nonetheless, it is possible to generate this hierarchical bottom to top
quark mass ratio, by uniformly subtracting two units of $U(1)_{A}$ charge
from all down fermion multiplets $\overline{\Psi }_{1}$, $\overline{\Psi }%
_{2}$ and $\overline{\Psi }_{3}$. The $Y^{d}$ matrix is then multiplied by
an overall factor of order $O(\epsilon ^{2})$. However, we do not make use
of this option here.}.

There is another important aspect of the model related with masses of quarks
and leptons which is worthy of special note. The $SU(7)$ GUT considered, as
well as the prototype $SU(5)$ model, predicts (see Eq.(\ref{15})) the
equality of the down quark and charged lepton masses at the grand
unification scale. However, due to the string scale unification in the $%
SU(7) $ model, the influence of the gravitational corrections on the Yukawa
couplings is much more important than in the prototype $SU(5)$ \cite{ellis}.
Actually, in general, one must include $SU(7)\otimes U(1)_{A}\otimes Z_{2}$
invariant corrections to the down fermion Yukawa couplings (\ref{15}, \ref{Y}%
, \ref{23g}) of the type

\begin{equation}
{\cal O}_{ij,\mbox{ }cor}^{down}\sim \frac{\epsilon ^{|\overline{q}%
_{i}+q_{j}|}}{M_{P}^{3}}(\overline{\Psi }_{i}\Sigma ^{2}\Psi _{j})(\overline{%
H}\varphi )  \label{cor}
\end{equation}
where $\overline{q}_{i}=Q_{A}(\overline{\Psi }_{i})$ and $q_{j}=Q_{A}(\Psi
_{j})$ are the $U(1)_{A}$ charges for the matter multiplets given in Eq.~(%
\ref{23g}). These gravitational corrections break the equality between the
down quark and charged lepton Yukawa couplings at the unification scale.
They give rise to effective dimensionless coupling constants $Y_{ij}^{\prime
}$ of order $\frac{<\Sigma >^{2}}{M_{P}^{2}}$ times the right hand side of
Eq.~(\ref{23g}). Thus, taking $<\Sigma >$ to be of order the string scale,
the corrections $Y_{ij}^{\prime }$ turn out to be of the same order as the
physical Yukawa couplings $Y_{ij}^{d}$ (e.g.~$Y_{33}^{\prime }\sim Y_{b,\tau
}\sim 0.01$) in the small $\tan \beta $ case. Thereby, under the
circumstances considered (string-scale unification plus small $\tan \beta $%
), gravitational corrections are expected not only to spoil the $SU(5)$
quark-lepton mass predictions for the $d$ and $s$ quarks, as argued in \cite
{ellis}, but also for the $b$ quark. At the same time these corrections will
not significantly disturb the hierarchical structure of the mass matrices (%
\ref{23f}) and (\ref{23g}), which is essentially protected by the $U(1)_{A}$
gauge symmetry. In the large $\tan \beta $ case the gravitational
corrections are quite small even for a string-scale unification and so the $%
b-\tau $ Yukawa unification, properly corrected by supersymmetric loop
contributions (see further discussion in Section 4), is actually predicted
as in the $SU(5)$ model.

\section{Gauge coupling unification}

We now consider gauge coupling unification in the above $SU(7)$ model. At
the low-energy scale, in addition to the three standard families of quarks
and leptons (and squarks and sleptons), there is just one family of PG
bosons (\ref{1}) and their superpartners which will modify the running of
the gauge and Yukawa couplings in the model. However, as we will see, the
contribution of the split adjoint moduli $\Sigma _{3}$ and $\Sigma _{8}$
taken at their natural scale $M_{P}^{2/3}M_{SUSY}^{1/3}$ (for earlier works
see \cite{yan}, \cite{su5tex}) turns out to be much more important.

We start with the reminder that, to one-loop order, gauge coupling
unification is given by the three renormalization group (RG) equations,
relating the values of the gauge couplings at the Z-peak $\alpha _{i}(M_{Z})$
$(i=1,2,3)$ and the common gauge coupling $\alpha _{U}$ at the unification
scale $M_U$ \cite{1}: 
\begin{equation}
\alpha _{i}^{-1}=\alpha _{U}^{-1}+\sum_{p}\frac{b_{i}^{p}}{2\pi }\ln\frac{%
M_{U}}{M_{p}}.  \label{8}
\end{equation}
Here $b_{i}^{p}$ are the three beta function coefficients, corresponding to
the $SU(7)$ subgroups $U(1)_{Y}$, $SU(2)_{W}$ and $SU(3)_{C}$ respectively,
for the particle labeled by $p$. The sum extends over all the contributing
particles in the model, and $M_{p}$ is the mass threshold at which each
decouples. All of the SM particles, and also the second electroweak doublet
of MSSM, are taken to be present at the starting scale $M_{Z}$. The next
contribution enters at the supersymmetric threshold, associated with the
decoupling of the supersymmetric particles at some single effective scale $%
M_{SUSY}$ \cite{3}. We propose to use relatively low values for this scale, $%
M_{SUSY} \sim M_{Z}$, so as to keep sparticle masses typically in the few
hundred GeV region. Furthermore, the PG states (\ref{1}) are also taken at a
sub-TeV scale. As to the heavy states, there are basic thresholds relating
to the adjoint moduli $\Sigma _{8}$ and $\Sigma _{3}$, with masses $M_{8}$
and $M_{3}$ respectively, and the thresholds caused by certain components in
the $SU(7)$ Higgs multiplets $H_{[AB]}$ and $\bar{H}^{[AB]}$,

\begin{equation}
H_{w7},\mbox{ }\overline{H}^{w7},H_{[cc^{\prime }]\mbox{ }},\overline{H}%
^{[cc^{\prime }]},  \label{R}
\end{equation}
developing masses of order $M_{*}$ (\ref{23b}), which we argued in Section
3.3 could lie somewhat lower than the GUT scale $M_{U}$. We refer to the
latter as the $H$-states. All other states in the Higgs, matter and gauge
multiplets, including the superheavy gauge bosons and their superpartners,
do not contribute to Eq.~(\ref{8}), for they are assumed to lie at the GUT
scale $M_{U}$, above which all particles fill complete $SU(7)$ multiplets.

Now, the $^{\prime \prime }$matching$^{\prime \prime }$ equation for the
gauge couplings (\ref{8}) reads as follows: 
\begin{eqnarray}
\lefteqn{12\alpha _{2}^{-1}-7\alpha _{3}^{-1}-5\alpha _{1}^{-1} \quad = } 
\nonumber \\
& &\frac{3}{2\pi }(-2\ln\frac{M_{U}}{M_{*}}+\ln\frac{M_{U}}{M_{3}} -7\ln%
\frac{M_{3}}{M_{8}}-\frac{19}{6}\ln\frac{M_{SUSY}}{M_{Z}})  \label{9}
\end{eqnarray}
This can be viewed as the basis for understanding qualitatively the
constraint on the value of $\alpha_3 = \alpha _{s}(M_{Z})$ from grand
unification, and its dependence on the present precision electroweak
measurements of $\alpha_1$ and $\alpha_2$ \cite{4} and all the thresholds
involved. One can see from Eq.~(\ref{9}) that $\alpha _{s}(M_Z)$ increases
with $\frac{M_{U}}{M_{3}}$ and decreases with $\frac{M_{U}}{M_{*}}$, $\frac{%
M_{SUSY}}{M_{Z}}$ and, especially, with $\frac{M_{3}}{M_{8}}$ (the term with
the largest coefficient in front of the logarithm). Paradoxically enough, in
the absence of all threshold effects ($M_{*}=M_{3}=M_{8}=M_{U}$ and $%
M_{SUSY}=M_{Z}$), Eq.~(\ref{9}) leads to a phenomenologically acceptable
value of $\alpha _{s}(M_{Z})$

\begin{equation}
\alpha _{s}=\frac{7\alpha _{em}}{15\sin^{2}\theta _{W}-3}=0.117,  \label{9aa}
\end{equation}
using the values $\alpha _{em}^{-1}=127.9$ and $\sin^{2}\theta_{W}=0.2313$
taken at $M_{Z}$ \cite{4}. Unfortunately, this value increases unacceptably
when 2-loop order corrections are included. In the standard $SU(5)$ case 
\cite{2}, with degenerate adjoint moduli $\Sigma _{3}$ and $\Sigma _{8}$ ($%
M_{3}=M_{8}$ at the GUT scale), an unacceptably high value of $\alpha_s$ is
obtained even if all the possible threshold effects are taken into account 
\cite{3,5,6}.

However, a drastically different unification picture appears when a generic
mass splitting between $\Sigma _{3}$ and $\Sigma _{8}$, which is a
consequence of the missing VEV inducing superpotential (\ref{2}), is taken
into account\footnote{%
This two-adjoint superpotential was recently applied to the $SU(5)$ model 
\cite{su5tex}. Although there can not be a missing VEV solution in the
framework of $SU(5)$, a generic mass-splitting between the adjoint moduli $%
\Sigma _{3}$ and $\Sigma _{8}$ appears ($\frac{M_{3}}{M_{8}}=4$) that leads
to a natural string-scale unification without any extension of the matter or
Higgs sectors in the model.}. Actually, after the adjoint scalars $\Sigma $
and $\Omega $ develop their VEVs (\ref{10}), the physical masses of the
surviving adjoint moduli $\Sigma _{3}$ and $\Sigma _{8}$ turn out to be
fixed. Diagonalization of the common $\Sigma $-$\Omega $ mass matrix

\begin{equation}
M_{ab}=[W_{ab}^{^{\prime \prime }}]_{<\Sigma >,<\Omega >}  \label{10a}
\end{equation}
(where the indices $a$ and $b$ stand for the corresponding components of $%
\Sigma $ and $\Omega $) leads to

\begin{equation}
M_{3}=\frac{14m}{3}~,~~~M_{8}=\frac{7m}{3}~,~~~\frac{M_{3}}{M_{8}}=2
\label{11}
\end{equation}
in contrast to $M_{3}/M_{8}=1$ in the standard $SU(5)$ model \cite{2}.

So, with the above observations, we are now ready to carry out the standard
two-loop analysis (with conversion from the $\overline{MS}$ scheme to the $%
\overline{DR}$ one included) \cite{1,12} for the gauge ($\alpha _{1}$, $%
\alpha_{2}$, $\alpha _{3}$) and Yukawa ($\alpha _{t}$, $\alpha _{b}$ and $%
\alpha_{\tau }$) coupling evolution. Here we are using the notation $\frac{%
Y_{t,b,\tau }^{2}}{4\pi }\equiv \alpha _{t,b,\tau }$ for the top- and
bottom-quarks and the tau-lepton. We include the standard supersymmetric
threshold corrections at low energies, taken at a single scale $%
M_{SUSY}=M_{Z} $, and those related with the PG states (\ref{1}) taken also
at a sub-TeV scale, namely at $300$ GeV. The heavy threshold corrections due
to the $H$ states (\ref{R}) with mass $M_{*}$ are taken close to the grand
unification scale $M_{U}$. As to the heavy $\Sigma $ adjoint moduli, their
masses were treated differently in the two parts of our calculation. When
making predictions of $\alpha _{s}(M_{Z})$ as a function of $M_{U}$ (see
Figure 1), they were varied from the MSSM$\;$unification point $M_{U}^{0}$ $%
\simeq 2\cdot 10^{16}$ GeV ($M_{3}=M_{U}^{0} $, $M_{8}=\frac{1}{2}M_{U}^{0}$%
) down to the intermediate value $m=O(10^{13})$ GeV \cite{yan}, thus pushing 
$M_{U}$ up to the string scale $M_{str}$. However, for the study of the
string-scale unification case $M_{U} $ $=M_{str}$ (see Tables 2 and 3),
these masses can be predicted and they, in fact, turned out to be at their
natural scale $M_{P}^{2/3}M_{SUSY}^{1/3}$ stemming from superstrings \cite
{yan}. The mass splitting between the weak triplet $\Sigma _{3}$ and the
colour octet $\Sigma _{8}$, which is fixed at the unification scale $M_{U}$
according to (\ref{11}), noticeably decreases as $M_{3}$ and $M_{8}$ run
down from $M_{U}$ to the lower energies according to their own two-loop RG
evolution. This effect was included in the analysis.

As to the Yukawa coupling evolution, we have considered both the cases of
small and large values of $\tan \beta $. The first case corresponds to $%
\alpha _{t}$ having a large enough value at the unification scale $M_{U}$ ($%
0.1>\alpha _{t}(M_{U})>0.01$) that it evolves towards its infrared fixed
point, while $\alpha _{b}(M_{U})$ and $\alpha _{\tau }(M_{U})$ are
significantlysmaller ($\alpha _{b,\tau }(M_{U})\lesssim 10^{-4}$). By
requiring the RG evolved value of $\alpha _{t}(m_{t})$ to reproduce the
observed value of the top quark pole mass, $m_{t}=175\pm 6$ GeV, the values
of $\tan \beta $ in Table 2 were determined. These values naturally satisfy
the usual RG infrared fixed point bound $\tan \beta >1.5$, which is just
consistent with the present experimental lower limit on the lightest MSSM
Higgs mass \cite{4}.

At the same time the direct $b-\tau $ unification prediction at the string
scale, $R_{b\tau }(M_{U})$ $=1$, for the bottom quark to tau lepton mass
ratio does not work well, when evolved down to low energies (see Table 2)
and compared to the experimental value \cite{4}

\begin{equation}
R_{b\tau }^{\exp }(M_{Z})=1.6\pm 0.2  \label{ex}
\end{equation}
However, as we argued in Section 3.6, the $SU(5)$-like equalities of down
quark and charged lepton masses, including $m_{b}(M_{U})=m_{\tau }(M_{U})$,
are highly unstable under gravity corrections for small $\tan \beta $ in the
string-scale $SU(7)$ GUT. So they are not expected to work well in our model.

The case of large $\tan \beta $, where all the couplings $\alpha _{t,b,\tau
} $ are relatively large and can approach a fixed-point, was found to have
acceptable solutions over the whole of the following range of starting
values: $0.01<\alpha _{t}(M_{U})<0.03$ and $0.001<\alpha _{b,\tau
}(M_{U})<0.3$. Here, however, the predictions for the top quark pole mass $%
m_{t}$, the ratio $R_{b\tau }(M_{Z})$ and $\tan \beta $ depend on detailed
information about the superparticle mass spectrum. Generally, in the large $%
\tan \beta $ case, large supersymmetric loop contributions to the bottom
quark mass $m_{b}$ are expected, which make the top mass prediction
uncertain as well. Since, fortunately, the direct SUSY loop contributions to
the $t$ quark and $\tau $ lepton masses are quite small \cite{13}, one can
adopt the following calculational strategy: calculate the value of $\tan
\beta $ using $\alpha _{\tau }(m_{\tau })$ from the RG equations and the
observed tau lepton mass $m_{\tau }=1.777$ GeV and use this value of $\tan
\beta $ to obtain the $m_{t}$ value. Further, for the given top mass $m_{t}$
one can calculate the size of the SUSY loop corrections to the bottom mass $%
\delta m_{b}$ required to bring the predicted $^{\prime \prime }$bare$%
^{\prime \prime }$ $R_{b\tau }^{0}(M_{Z})$ values given in Table 3 into
agreement with experiment

\begin{equation}
R_{b\tau }(M_{Z})=R_{b\tau }^{0}(M_{Z})[1+\Delta ]=R_{b\tau }^{\exp }(M_{Z})
\label{th}
\end{equation}
(see $R_{b\tau }^{\exp }(M_{Z})$ above). Interestingly, as one can see from
Table 3, the required values of $\Delta $ turn out to be in the range $%
\Delta =-0.3\div -0.1$. The SUSY loop corrections in this range are readily
obtained for most of SUSY parameter space \cite{13} with the values of $%
\alpha _{s}(M_{Z})$ and $m_{t}$ in Table 3. So, presently, the bottom-tau
unification prediction on its own does not seem to be a critical test of the 
$SU(7)$ model for either large or small $\tan \beta $.

Our results, obtained by numerical integration of all the RG equations
listed above, are summarized in Figures (1, 2) and Tables (2, 3). The
predicted $\alpha _{s}(M_{Z})$ values are in a good agreement with the world
average value (see Figure 1), in sharp contrast to the standard SUSY $SU(5)$
model predictions taken under the same conditions . Examples of string-scale
unification are presented in Tables 2 and 3 for the small and large $\tan
\beta $ cases, respectively. It is worth noting that the large $\tan \beta $
examples actually include $^{\prime \prime }$medium$^{\prime \prime }$
values of $\tan \beta $ down to $\tan \beta \approx 20$, in contrast to the
prototype SU(5) model. Remarkably, string-scale unification appears to work
both in the case of a large bottom Yukawa coupling constant (the first line
in Table 3), and in the case of top-bottom unification with a small common
Yukawa coupling constant (the third line).

Thus we have seen how the missing VEV generating superpotential $W_{A}$ (\ref
{2}) opens the way to a natural string-scale grand unification in
supersymmetric $SU(7)$, prescribed at low energies by the gauge coupling
values and the standard MSSM particle content plus one family of PG states (%
\ref{1}). This is explicitly demonstrated in Figure 2.

At the same time, due to the reflection symmetry ($\Sigma \to -\Sigma $, $%
\Omega \to \Omega $) of the superpotential $W_{A}$, the Planck scale
inherited smearing operators, which induce a $\Sigma $ dependence into the
kinetic terms of the SM gauge bosons, must have dimension 6 and higher,

\begin{equation}
\delta L=\frac{c}{M_{P}^{2}}~Tr(GG\Sigma ^{2})+...  \label{RR}
\end{equation}
Here $G$ is the gauge field-strength matrix of the $SU(7)$ model and $c$ is
some dimensionless constant of order $1$. Thus, the influence of
gravitational corrections on our gauge coupling predictions seems to be
negligible, in contrast to the standard $SU(5)$ model predictions which can
largely be smeared out by the dimension 5 operator $\frac{c^{\prime}}{M_{P}}%
~Tr(GG\Sigma)$ \cite{hall}.

\section{Conclusions}

We have shown that a missing VEV vacuum configuration, which solves the
doublet-triplet splitting problem and ensures the survival of the MSSM at
low energies, only emerges in extended $SU(N)$ SUSY GUTs with $N \ge 7$.
Furthermore, a realistic supersymmetric $SU(7)$ model was constructed which
provides answers to many questions presently facing the prototype SUSY $%
SU(5) $ model: the doublet-triplet splitting problem, string scale
unification, the hierarchy of baryon vs lepton number violation, quark and
lepton (particularly neutrino) masses and mixings etc.

With the chosen assignment of matter and Higgs superfields in our $SU(7)$
model, the situation at low energies looks as if one had just the prototype $%
SU(5)$ as a starting symmetry, except that one family of PG states of type (%
\ref{1}) appears when a missing VEV vacuum configuration develops in $SU(7)$%
. Apart from this exception, all other extra $SU(7)$ inherited states in
matter and Higgs multiplets acquire GUT scale masses during symmetry
breaking, thus completely decoupling from low-energy physics. Another
distinctive feature of our model concerns the relatively low mass scale of
the adjoint moduli $\Sigma _{0,}$ $\Sigma _{3}$ and $\Sigma _{8}$ surviving
after the $SU(7)$ breaks and, more importantly, their mass-splitting which
inevitably appears in the missing VEV generating superpotential (\ref{2}).
The threshold corrections due to these states lead to a very different
unification picture in $SU(7)$. String scale unification is obtained with a
QCD coupling constant in agreement with the world average value, $%
\alpha_s(M_Z)=0.119 \pm 0.002$, for both small and large $\tan \beta $.

There is no fundamental reason for exact $R$-parity ($RP$) symmetry in the
framework of supersymmetric GUTs, where not only fermions but also their
scalar superpartners are the natural carriers of lepton and baryon numbers.
Accordingly, we constructed a general ($RP$-violating) superpotential where
the effective lepton number violating couplings immediately evolve from the
GUT scale. However the baryon number non-conserving ones are safely
projected out by the missing VEV vacuum configuration which breaks the $%
SU(7) $ symmetry down to that of the MSSM. At the next stage when SUSY
breaks, the radiative corrections shift the missing VEV components to some
nonzero values of order $M_{SUSY}$, thereby inducing the ordinary Higgs
doublet mass, on the one hand, and tiny baryon number violation, on the
other. So, a missing VEV solution to the gauge hierarchy problem leads in
fact to a similar hierarchy of baryon vs lepton number violation. Finally,
an additional anomalous $U(1)_{A}$ symmetry introduced into the model purely
as a missing VEV protecting symmetry was found to naturally act also as a
family symmetry, which can lead to the observed pattern of quark and lepton
masses and mixings.

To conclude the most crucial prediction of the presented $SU(7)$ model must
surely be the very existence of the PG states and their superpartners of
type (\ref{1}). While at present not producing any unacceptable experimental
predictions, they may include a few very long-lived states, depending on the
details of their mixing pattern with the ordinary MSSM Higgs states. The
colour-triplet states among them may gather together with ordinary quarks to
give a new series of heavy hadrons (mesons and baryons), the lightest of
which could be almost stable \cite{gift}. These PG states would clearly
strongly influence particle phenomenology at the TeV scale.

\section*{Acknowledgments}

We would like to thank many of our colleagues, especially Riccardo Barbieri,
Zurab Berezhiani, Grahame Blair, Gia Dvali, Mike Green, David Hutchcroft,
Gordon Moorhouse, S. Randjbar-Daemi, Alexei Smirnov and David Sutherland,
for stimulating dicussions and useful remarks. Financial support by INTAS
Grants No. RFBR 95-567 and 96-155 , and a Joint Project grant from the Royal
Society are also gratefully acknowledged.

\section*{Figure captions\bigskip}

{\bf Figure 1:}~ The $SU(7)$ predictions of $\alpha _{s}(M_{Z})$ as a
function of the grand unification scale $M_{U}$. Curves are shown for small $%
\tan \beta $ values (the dotted line) with top-Yukawa coupling $\alpha
_{t}(M_{U})=0.1$ and the $H$-state mass threshold $M_{*}=10^{16}$ GeV, and
for large $\tan \beta $ values (the solid line) corresponding to $\alpha
_{t}(M_{U})=0.02$, $\alpha _{b}(M_{U})=0.1$ and $M_{*}=10^{16.8}$ GeV. The
unification mass $M_{U}$ varies from the MSSM unification point ($%
M_{U}^{0}=10^{16.3}$ GeV) to the string scale ($M_{U}=M_{str}=10^{17.8}$ GeV
for the Kac-Moody level $k=2$), while the proton decay inducing
colour-triplet mass is assumed to be at the unification scale in all cases.
The all-shaded areas on the left are generally disallowed by the present
bound \cite{4} on nucleon stability for both cases, (small $\tan \beta $,
dark) and (large $\tan \beta $, light) respectively.\medskip

\noindent {\bf Figure 2:}~ The unification of the gauge coupling constants $%
\alpha _{1}$, $\alpha _{2}$ and $\alpha _{3}$ at the string scale is shown.
The breaks on the curves correspond to the heavy thresholds associated with
the adjoint moduli $\Sigma _{3}$ and $\Sigma _{8}$ (on the left) and the $H$
states (on the right).

\section*{Tables}

\bigskip

{\bf Table 1: }The $U(1)_{A}$ ($e^{iQ_{A}\theta }$) and $Z_{2}$ ($%
e^{in_{Z}\pi }$) charges, which allow just the couplings appearing in the
total superpotential $W_{T}$ (\ref{wwww}).

\medskip

\[
\begin{tabular}{||l|lllllllllllllll||}
\hline\hline
${\cal F}$ & $\Psi $ & $\overline{\Psi }$ & $\overline{\xi }$ & $\overline{%
\zeta }$ & $H$ & $\overline{H}$ & $\Sigma $ & $\Omega $ & $\varphi $ & $%
\overline{\varphi }$ & $\eta $ & $\overline{\eta }$ & $\Phi $ & $\overline{%
\Phi }$ & $S$ \\ 
&  &  &  &  &  &  &  &  &  &  &  &  &  &  &  \\ 
$Q_{A}$ & $0$ & $0$ & $-3$ & $-1$ & $-2$ & $2$ & $0$ & $0$ & $-2$ & $3$ & $2$
& $1$ & $-5$ & $4$ & $1$ \\ 
$n_{Z}$ & $1$ & $0$ & $0$ & $0$ & $0$ & $1$ & $1$ & $0$ & $0$ & $1$ & $0$ & $%
1$ & $0$ & $0$ & $0$ \\ \hline\hline
\end{tabular}
\]

\bigskip

{\bf Table 2: }The 2-loop order string-scale unification for the small $\tan
\beta $ case is presented. All masses are given in GeV. The top quark Yukawa
coupling $\alpha _{t}(M_{U})$ and $H$-state mass threshold $M_{*}(M_{U})$
are taken as basic string scale input parameters ($M_{U}=M_{str}=\sqrt{8\pi
\alpha _{U}}\cdot 5.27\cdot 10^{17}$ for the Kac-Moody level $k=2$), while
low scale input parameters include $\alpha _{em}^{-1}=127.9\pm 0.1$, $\sin
^{2}\theta _{W}=0.2313\pm 0.0002$ and the top quark pole mass $m_{t}=175\pm
6 $. The values of $\alpha _{s}(M_{Z})$, the $SU(7)$ unified coupling
constant $\alpha _{U}$, $\tan \beta $, the $^{\prime \prime }$bare$^{\prime
\prime }$ (uncorrected by gravitational contributions) bottom-tau mass ratio 
$R_{b\tau }^{0}(M_{Z})$ and the adjoint moduli triplet $\Sigma _{3}$ mass $%
M_{\Sigma }(M_{U})$ (the adjoint octet $\Sigma _{8}$ mass is $\frac{1}{2}
M_{\Sigma }(M_{U})$) are then predicted.

\bigskip

\begin{tabular}{||ccccccc||}
\hline\hline
$\alpha _{t}(M_{U})$ & $M_{*}(M_{U})$ & $M_{\Sigma }(M_{U})$ & $\alpha _{U}$
& $\alpha _{s}(M_{Z})$ & $\tan\beta $ & $R_{b\tau }^{0}(M_{Z})$ \\ 
&  &  &  &  &  &  \\ 
$0.1$ & $10^{16}$ & $10^{13.3}$ & $0.080$ & $0.120$ & $1.6$ & $2.2$ \\ 
$0.08$ & $10^{15.8}$ & $10^{13.1}$ & $0.082$ & $0.119$ & $1.6$ & $2.3$ \\ 
$0.04$ & $10^{15.4}$ & $10^{12.9}$ & $0.087$ & $0.119$ & $1.8$ & $2.4$ \\ 
$0.02$ & $10^{15.2}$ & $10^{12.8}$ & $0.090$ & $0.119$ & $2.2$ & $2.6$ \\ 
$0.01$ & $10^{15}$ & $10^{12.8}$ & $0.092$ & $0.119$ & $4.8$ & $2.7$ \\ 
\hline\hline
\end{tabular}

\bigskip

\bigskip

{\bf Table 3: }The 2-loop order string-scale unification for the large $\tan
\beta $ case is presented. All masses are given in GeV. The Yukawa couplings 
$\alpha _{t,b,\tau }(M_{U})$ and $H$-state mass threshold $M_{*}(M_{U})$ are
taken as basic string scale input parameters ($M_{U}=M_{str}=\sqrt{8\pi
\alpha _{U}}\cdot 5.27\cdot 10^{17}$ for the Kac-Moody level $k=2$), while
low scale input parameters include $\alpha _{em}^{-1}=127.9\pm 0.1$ and $%
\sin ^{2}\theta _{W}=0.2313\pm 0.0002$ . The values of $\alpha _{s}(M_{Z})$,
the $SU(7)$ unified coupling constant $\alpha _{U}$, $\tan \beta $, the top
quark pole mass $m_{t}$, the $^{\prime \prime }$bare$^{\prime \prime }$
(uncorrected by SUSY loop contributions) bottom-tau mass ratio $R_{b\tau
}^{0}(M_{Z})$ and the adjoint moduli triplet $\Sigma _{3}$ mass $M_{\Sigma
}(M_{U})$ (the adjoint octet $\Sigma _{8}$ mass is $\frac{1}{2}M_{\Sigma
}(M_{U})$) are then predicted.

\bigskip

\begin{tabular}{||ccccccccc||}
\hline\hline
$\alpha _{t}(M_{U})$ & $\alpha _{b,\tau }(M_{U})$ & $M_{*}(M_{U})$ & $%
M_{\Sigma }(M_{U})$ & $\alpha _{U}$ & $\alpha _{s}(M_{Z})$ & $m_{t}$ & $\tan
\beta $ & $R_{b\tau }^{0}(M_{Z})$ \\ 
&  &  &  &  &  &  &  &  \\ 
0.025 & 0.3 & 10$^{17.8}$ & 10$^{14}$ & 0.066 & 0.120 & 176 & 55.1 & 1.9 \\ 
0.02 & 0.1 & 10$^{16.8}$ & 10$^{13.5}$ & 0.073 & 0.119 & 175 & 52.8 & 1.9 \\ 
0.01 & 0.01 & 10$^{15.2}$ & 10$^{12.7}$ & 0.089 & 0.118 & 172 & 38.9 & 2.3
\\ 
0.01 & 0.004 & 10$^{15.2}$ & 10$^{12.8}$ & 0.089 & 0.119 & 175 & 30.3 & 2.4
\\ 
0.01 & 0.001 & 10$^{15.2}$ & 10$^{12.9}$ & 0.090 & 0.120 & 179 & 17.9 & 2.6
\\ \hline\hline
\end{tabular}

\end{document}